# A 0.08 pJ/bit 56 GBaud Monolithic Optical Receiver Front End for IMDD Photonic Links

**Robert P. Pesch**, ***Graduate Student Member, IEEE***, **Arjun Khurana**, ***Graduate Student Member, IEEE***, **Joshua J. Wong**, ***Graduate Student Member, IEEE***, **Joel Slaby**, ***Graduate Student Member, IEEE***, and **Stephen E. Ralph**, ***Senior Member, IEEE***

School of Electrical and Computer Engineering, Georgia Institute of Technology, Atlanta, GA 30332

Corresponding author: Stephen E. Ralph (e-mail: Stephen.ralph@ece.gatech.edu).

This material is based upon work supported by the National Science Foundation Graduate Research Fellowship under Grant No. DGE-2039655 and National Science Foundation “EPICA”, Grant No. 252808.

**ABSTRACT** We present the design, fabrication, and measurement of a monolithically integrated optical receiver analog front end to achieve low power operation while supporting 56 GBaud intensity modulated direct detect transceivers. Transistor unit cell layout optimization is utilized to minimize parasitics while enabling wide analog bandwidth and reduced input referred noise. The circuit, fabricated on the GlobalFoundries Fotonix$^{TM}$ monolithic silicon photonics platform, is characterized by its DC operation, AC operation, noise characteristics, and time domain behavior. The final design was validated by on-off keyed and PAM-4 electrical eye diagram measurements to 64 GBaud, consuming 9.22 mW of power from a 1.2 V supply. The measured circuit achieves 61.5 dBΩ transimpedance gain and 29.5 GHz electrical BW with less than 737 nA RMS integrated input referred noise current and 0.08 pJ/bit.

## I. INTRODUCTION

High capacity optical links are increasingly deployed in systems where baud rate scaling is constrained not only by bandwidth (BW), but also by power consumption, thermal management, and integration complexity [1-2]. Two notable examples are artificial intelligence (AI) datacenters and aerospace platforms. Power and thermal considerations have become dominant barriers to scaling of AI networks creating strong demand for dense, high-throughput optical interconnects with low energy per bit [1]. Similarly, aerospace systems, including satellites, require high-capacity data distribution under stringent size, weight, power, and cost (SWaP-C) metrics [2]. In both domains, intensity modulated direct detect (IMDD) photonic links remain valuable due to their reach, low latency, and power efficiency [3,4].

Despite these advantages, power consumption, heat dissipation, and packaging complexity remain challenging for deployed optical transceivers. In datacenter optical modules, optical receiver power directly contributes to module heat density and system level cooling demand. This problem is exacerbated on satellite platforms where total payload power is limited, and heat removal is challenging due to the absence of convective cooling. Active thermal control solutions, such as thermo-electric coolers (TECs), stabilize component temperatures at the cost of additional system complexity and increased power consumption; therefore, reducing optical receiver power is beneficial not only for energy efficiency but also thermal management and scalability.

Here, we address these concerns through the design of a monolithically integrated optical receiver analog front end (AFE). Monolithic integration significantly reduces both packaging parasitics and system complexity, compensating for the reduced transit frequency ($f_t$) relative to conventional radio frequency (RF) CMOS platforms. The proposed receiver integrates a germanium photodiode, a modified active voltage-current feedback (AVCF) transimpedance amplifier (TIA), cascaded broadband post amplifier stages, and an output buffer to achieve record low power consumption, low noise, and wideband operation. These characteristics make the receiver eligible for power constrained IMDD optical links in AI datacenter and aerospace applications.

The work is organized into four additional sections. Section II describes the system-level receiver requirements and the circuit-level design of the analog front-end blocks. Section III discusses the physical design, transistor layout, and parasitic-aware optimization technique. Section IV presents the experimental characterization, including DC power, AC transimpedance gain, noise, and time-domain eye-diagram measurements, and compares the results with the reported literature. Pure electrical characterization is highlighted to emphasize the design considerations laid out in Sections II and III. Finally, Section V identifies future challenges and discusses ongoing work.

## II. Receiver Design

A conventional AFE consists of the following, Fig. 1:

### 1) PHOTIODE (PD)

The photodiode is characterized by a capacitance, an opto-electronic bandwidth, and a responsivity operating under a reverse bias. For PAM-4 operation, optical power is limited to ensure operation in the linear regime.

### 2) TRANSIMPEDANCE AMPLIFIER (TIA)

The TIA presents a low input impedance to the photodiode to minimize the associated RC time constant and dominates the noise characteristics of the total AFE.

### 3) POST AMPLIFIER

Additional gain is often necessary to reach the required voltage swing for subsequent digital circuitry. The post amplifier is implemented in several cascaded low-gain stages to maintain bandwidth.

### 4) OUTPUT BUFFER (OB)

An output buffer is optimized to efficiently drive the load ($Z_L$) by both sourcing sufficient drive current and impedance matching to maximize power transfer while minimizing back reflections.

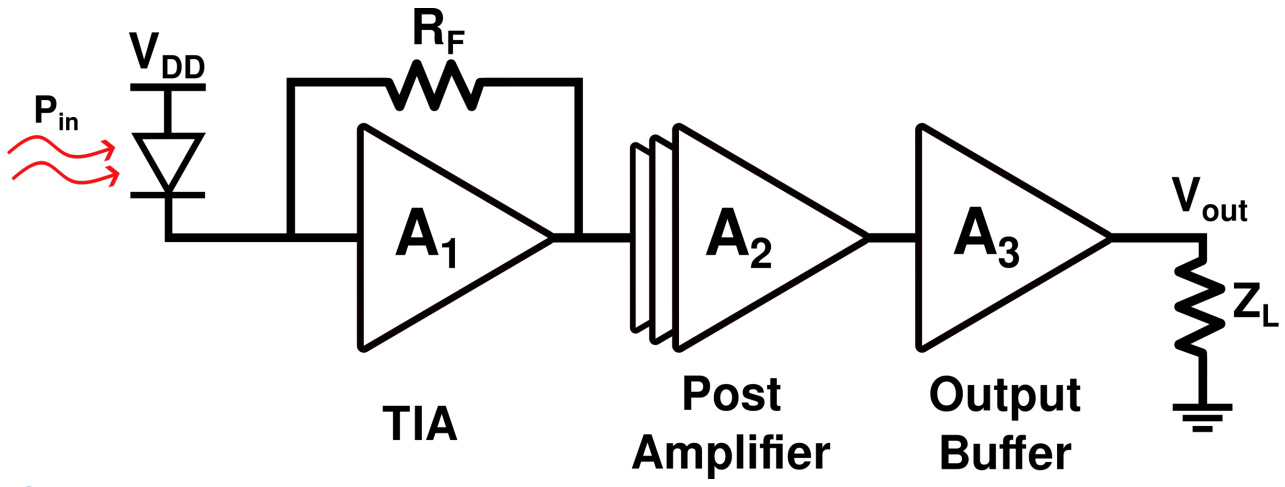

**FIGURE 1.** Block diagram of conventional photonic receiver AFE with input photodiode, TIA for current to voltage conversion, post amplifier for voltage gain, and output buffer for impedance matching.

We note that many AFEs deployed in ultra-high data rate systems also include a continuous time linear equalizer (CTLE) between the TIA and the post amplifier [6]. This maximizes the net bandwidth of the receiver while minimizing the input referred noise current (IRNC). Here, we start with designing an AFE that prioritizes the low-power requirement, noting that a CTLE can be included later.

The specific circuit architectures for the TIA, post amplifier, and OB are dependent on the target baud rate. For the GF Fotonix™ platform, 56 GBaud is feasible requiring a BW of at least 28 GHz.

The total transimpedance gain of the receiver is determined from the expected input optical power and required output voltage swing. For CMOS circuitry operating with ~1 V supply, the desired output swing is on the order of several hundred milli-Volts. Whereas anticipated integrated photonic systems optical powers are on the order of hundreds of micro-Watts (-10dBm). Therefore, we seek a transimpedance gain on the order of 60 dBΩ with an assumed near unity photodiode responsivity. This metric includes both the TIA transimpedance gain as well as the subsequent post amplifier and OB voltage gain. The optimization and tradeoffs between allocating gain to each of these stages is detailed subsequently.

Mitigating noise in the targeted photonic links requires addressing noise sources in both the optical and electrical domains. Here, we focus on minimizing the electrical noise from the entire AFE given the gain and BW targets are satisfied. In the context of optical receiver front ends, noise is represented as the input referred noise current (IRNC). The IRNC is found by combining all the effects of every noise source in the entire AFE, often including the photodiode dark current shot noise, and referring this to the input node as an additional parallel current source Fig. 2. Reported AFEs with transimpedance gains on the order of 60 dBΩ or BW in excess of 30 GHz achieve IRNC PSDs of tens of pA/Hz$^{1/2}$ , so here we an IRNC of less than 10 pA/Hz$^{1/2}$[7].

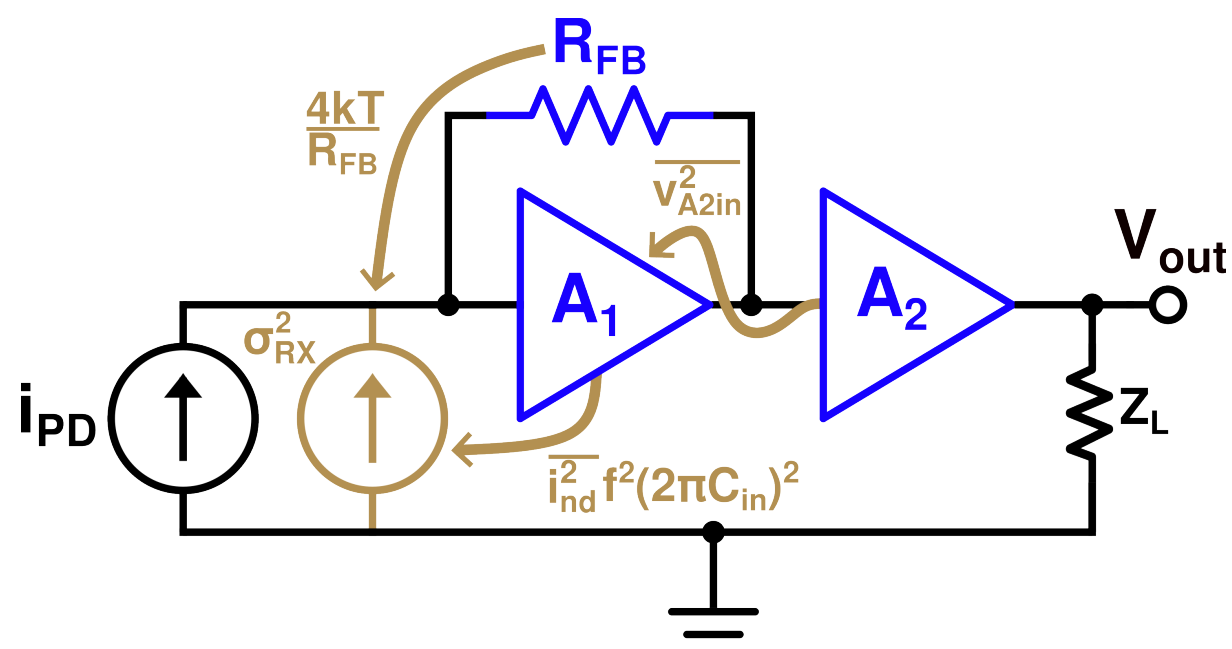

**FIGURE 2.** Simplified schematic of AFE driven by output current of the photodiode ($i_{PD}$), highlighting the various noise contributions from the TIA (consisting of voltage amplifier $A_1$ and feedback resistor $R_{FB}$), as well as the subsequent post amplifier and OB (captured together as an effective voltage amplifier $A_2$). The input referred noise current PSD is composed of the feedback resistor thermal current noise and the noise of both amplifier stages referred to the input as a Norton equivalent current source. By making the gain of the TIA stage large, we can neglect the noise contributions from subsequent stages (Section II.B).

Finally, total power consumption is the most critical design target for this AFE and will be minimized subject to the gain, bandwidth, and noise targets. Based upon desired sub pJ/bit operation and the knowledge that the Tx typically requires more power total power consumption of less than 10 mW is desired.

### *A. PHOTODIODE*

Photodiodes within a silicon photonic platform are typically comprised of germanium and have excellent responsivity from O-band (1300nm) to L band (1625nm). The photodiode in GlobalFoundries Fotonix™ platform [8] has a responsivity of ~0.9 A/W, an opto-electronic bandwidth of 40 GHz, and a dark current of less than 40 nA. The PD operates under a reverse bias condition and is directly coupled to the subsequent TIA stage.

## B. TRANSIMPEDANCE AMPLIFIER

Many TIA topologies have been explored with applications in fields beyond photonics (e.g. biomedical engineering, microelectromechanical systems, etc.) [9-11], all with inherent gain, noise, and 3-dB bandwidth trade-offs. We seek a topology that balances BW and IRNC while minimizing power. The gain of this stage needs to be sufficient to reduce the impact of the noise from the post amplifier and OB but not be so large to overly reduce the BW. For a total transimpedance gain of 60 dBΩ, a 45-50 dBΩ is a sufficient starting point.

While conventional TIAs utilize a high gain amplifier with resistive feedback, other feedback topologies offer superior bandwidth and power tradeoffs [9]. Additionally, topologies with inductors will not be explored due to their large footprint. Here we demonstrate a new modified version of the inverter-based active voltage-current feedback (AVCF) TIA topology, Fig. 3a [12,13,26] due to its potential for low-power high speed operation. Our AVCF implementation includes degeneration resistors $R_{SN}$ and $R_{SP}$ without the additional feedback resistors and voltage amplifiers used in [13]. These changes improve the tunability of the architecture without adding unnecessary power consumption and design complexity.

To clarify the design tradeoffs motivating the modified AVCF topology, it is compared against a conventional resistive-feedback TIA. In the resistive-feedback TIA, the low-frequency transimpedance is set primarily by the feedback resistance, while the input pole, often the dominant pole, is determined by the effective input resistance and the input capacitance. Increasing the feedback resistance improves gain and reduces the IRNC but also reduces bandwidth and can require additional amplifier gain and bias current to preserve stability. Therefore, high-gain, wide bandwidth, and low-noise are not independently tunable. In contrast, the proposed modified AVCF TIA separates these tradeoffs more effectively. The input impedance is reduced through active current feedback rather than only by increasing the forward amplifier gain. The input pole can then be pushed to higher frequency at a lower bias current. The feedback resistor and source-degeneration resistors then provide additional control of gain and damping. This allows the TIA core to achieve sufficient front end gain to suppress subsequent post amplifier and output buffer noise while maintaining a wide input bandwidth. These mitigate the need for power-hungry high-gain broadband post amplifier stages and large area peaking inductors.

The associated small-signal model, Fig. 3b, was analyzed to derive the frequency response of the transimpedance gain, $Z_T(s)$. This function can be modelled as a second order low-pass filter (LPF), with a resonant frequency and a damping factor. $Z_T(s)$ can further be broken down into the input impedance frequency response $Z_{in}(s)$ and the voltage gain $A_v$. The full low-frequency input impedance is the parallel combination of the impedance looking into the drains of $M_{N1}$ and $M_{P1}$.

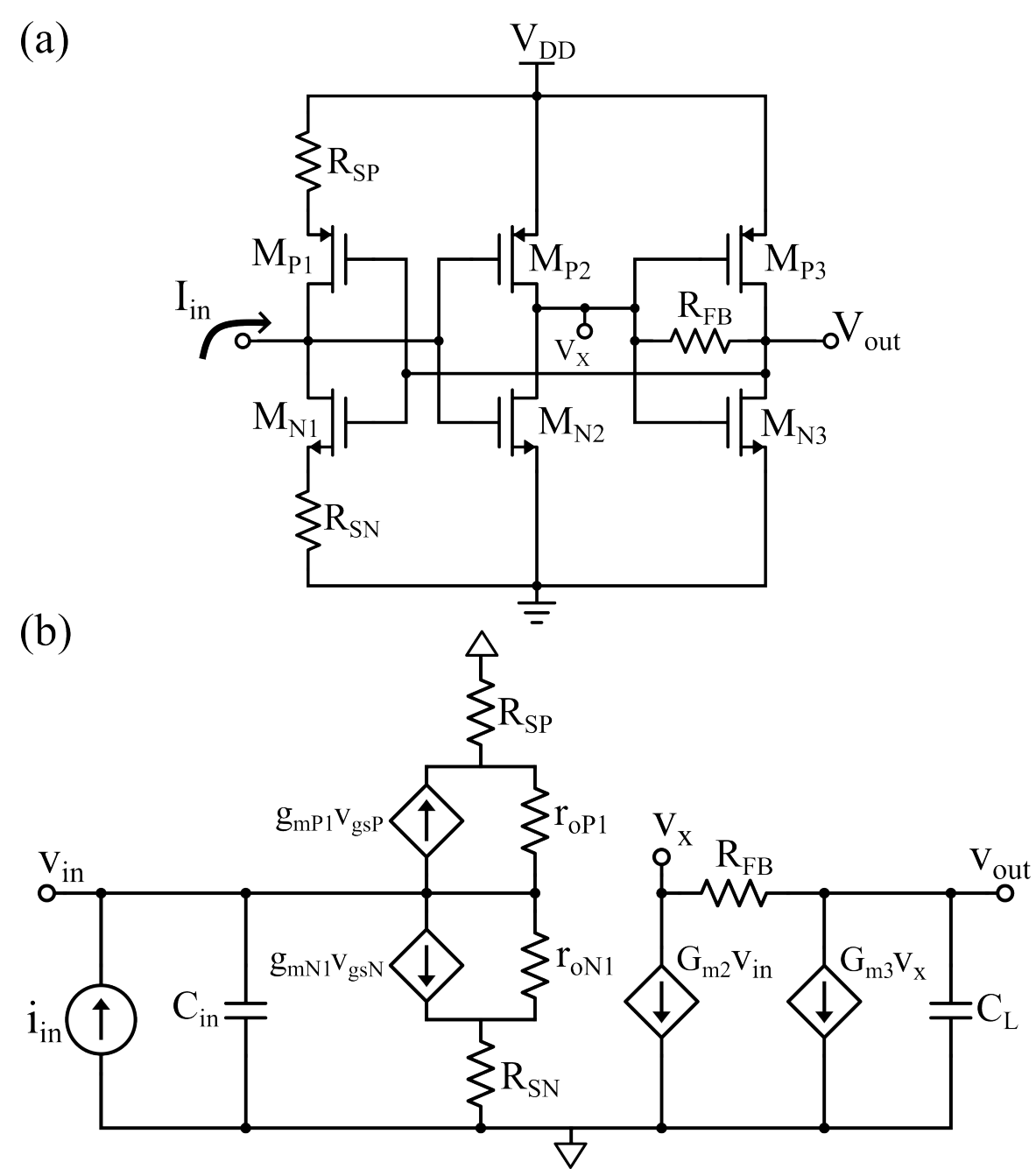

**FIGURE 3.** Schematic diagram of modified AVCF TIA topology (a) and associated small-signal model (b) utilized for circuit analysis. $G_{M2}$ and $G_{M3}$ represent the combined transconductances of the NMOS and PMOS devices in the second and third stages respectively.

$$Z_{in}(0) = \frac{r_{oN1}(1 + g_{mN1}R_{SN}) + R_{SN}}{1 + A_v g_{mN1} r_{oN1}} \parallel \frac{r_{oP1}(1 + g_{mP1}R_{SP}) + R_{SP}}{1 + A_v g_{mP1} r_{oP1}} \tag{1}$$

Where $r_{oN1}$ ($r_{oP1}$) and $g_{mN1}$ ($g_{mP1}$) represent the small-signal output resistance and transconductance for the transistor $M_{N1}$ ($M_{P1}$) as illustrated in the equivalent small-signal model, Fig. 3b. The MOSFET small-signal output resistances are included for the first stage to highlight how the $R_S$ resistors modify the input impedance, while the second and third stage resistances can be excluded to simplify the intuition behind design trade-offs. From Fig. 3b, $A_v$ is calculated as:

$$A_v = G_{m2}R_{FB} - G_{m2}/G_{m3} \tag{2}$$

Here, $G_{m2}$ and $G_{m3}$ are the effective transconductances of the second and third stages respectively, where $G_{mi} = g_{mNi} + g_{mPi}$. With proper sizing and biasing, the respective small-signal parameters of each stage's NMOS and PMOS can be reasonably approximated as equal. That is, $g_{mN1} = g_{mP1} = g_{m1}$ and $r_{oN1} = r_{oP1} = r_{o1}$. Setting $R_{SN} = R_{SP} = R_S$, for symmetry, the input impedance can be re-written as:

$$Z_{in}(0) = \frac{r_{o1}(1 + g_{m1}R_S) + R_S}{2(1 + A_v g_{m1} r_{o1})} \tag{3}$$

Since, $v_{in} = Z_{in}(s) \times i_{in}$, the full transimpedance transfer function of the TIA core can be expressed as $Z_T(s) = A_v \times Z_{in}(s)$:

$$Z_T(s) = \frac{1}{C_{in}} \cdot \frac{1}{s + \frac{1}{Z_{in}(0)C_{in}}} \times \frac{1}{C_L} \cdot \frac{G_{m2}G_{m3}R_{FB} - G_{m2}}{s + \frac{G_{m3}}{C_L}} \quad (4a)$$

$$Z_T(s) = \frac{1}{C_{in}C_L} \cdot \frac{G_{m2}G_{m3}R_{FB} - G_{m2}}{s^2 + 2\xi\omega_n s + \omega_n^2} \quad (4b)$$

$$\omega_n = \sqrt{\frac{1}{C_{in}C_L} \cdot \frac{2g_{m1}(G_{m2}G_{m3}R_{FB} - G_{m2})}{1 + g_{m1}R_S}} \quad (5)$$

$$\xi = \sqrt{\frac{C_{in}}{8C_L} \cdot \frac{G_{m3}(1 + g_{m1}R_S)}{g_{m1}(G_{m2}R_{FB} - \frac{G_{m2}}{G_{m3}})}} \quad (6)$$

Where $C_{in}$ is the total effective capacitance at the input node, $C_L$ is the output capacitance, $\omega_n$ is the resonant frequency of the second order transfer function and $\xi$ is a damping coefficient. From (4a) the poles of $Z_{in}(s)$ and $A_v(s)$ can be identified as $p_1 = -1/(Z_{in}(0)C_{in})$ and $p_2 = -G_{m3}/C_L$ respectively. Equations (5) and (6) assume that the magnitude of $p_2$ is much larger than $p_1$. This assumption is valid due to the DC input impedance being on the order of $1/G_{m3}$. Moreover, the monolithic implementation ensures that the load capacitance is larger than the input capacitance. This is because the input capacitance only consists of the small photodiode contribution (~10fF) and the lumped transistor parasitic capacitances, near that of the PD for optimal noise characteristics, while the output capacitance is dominated by the large output pad contribution (~60 fF).

Another consideration regarding the frequency response of this circuit is the Miller effect. Since each individual stage is an inverting amplifier, capacitive feedback between the input and output (primarily the sum of the NMOS and PMOS gate-source capacitances) will be effectively amplified [14], reducing the 3-dB bandwidth of the circuit. However, since the intrinsic gain and gate-source capacitances of the final sized $M_{N1}$ and $M_{P1}$ are small (less than 7 V/V and several hundred attofarads respectively) the amplified Miller capacitance is less than a few femtofarads and does not noticeably inhibit performance. Other works have explored buffering the output voltage before feeding it back to the first stage of the AVCF structure [13] to reduce the Miller effect. Here it was omitted to minimize power consumption while offering no appreciable performance benefits

Next, the small-signal model is analyzed for the circuit's noise behavior. Following the results from [12] the IRNC PSD can be written as:

$$\overline{i_{in}^2} = 4kT\Gamma G_{m1} + \frac{8kT}{R_S} + \frac{4kT\Gamma C_{in}^2\omega^2}{G_{m2}} + \frac{4kT\Gamma C_{in}^2\omega^2}{G_{m2}^2 R_{FB}} + \frac{4kT\Gamma C_{in}^2\omega^2}{G_{m2}^2 G_{m3} R_{FB}^2} \quad (7)$$

Where $k$ is Boltzmann's constant, T is the temperature in Kelvin, $\Gamma$ is Ogawa's excess noise factor [15], and $\omega$ is radial frequency. This expression omits transistor flicker noise due to the anticipated high frequency operation. Instead, (7) is dominated by both a white noise component, terms one and two, and a violet noise component, terms three through five. For similar TIAs designed with sub 10 GHz bandwidths, the violet component is often negligible; however, for high data rate photonic applications, the rapid increase of the noise PSD after this point leads to a measurable increase in the RMS output noise voltage.

For practical design considerations, the tuning resistors of the TIA core enable both high gain and ultra-wide BW operation. By increasing the feedback resistance $R_{FB}$, the gain and BW both increase, and the damping factor decreases. If $\xi$ becomes too small, the resulting loop phase margin becomes less than 45° leading to ringing in the step response. However, by decreasing $\xi$ without entering the unstable regime, the resulting peaking in the frequency domain can be utilized with a subsequent post-amplifier and output buffer to result in a flat wide-band response (Fig. 5). In this design, we chose a moderate value of $R_{FB}$ which exhibits slight peaking ($\xi = 0.55$) in the core response but not the cascaded response, ensuring stable operation. Additionally, the series resistors $R_{SN}$ and $R_{SP}$ give the designer flexibility to tune the gain and bandwidth of the TIA without affecting the stability of the frequency response. However, increasing these resistances will also increase the PSD of the resistor based thermal noise in the first stage of the TIA core, increasing the IRNC.

Transistor sizing and biasing is next addressed. Because each stage is symmetric about the respective input and output stages, the DC level of each node will be ~$V_{DD}/2$, if the transistors are sized properly. The input stage is sized small but not minimally, to reduce the parasitic capacitance between the output and input nodes while keeping the input impedance small compared to the photodiode output impedance. As for the second and third stages, an exponential horn structure [16] is employed to size the stages relative to one another. No external bias circuitry is required with this architecture, reducing power that would be consumed by current mirror or bandgap references. With the tuning resistor values chosen and the transistors sized, the TIA core is complete.

### *C. POST AMPLIFIER AND OUTPUT BUFFER*

A simple cascade of several Cherry-Hooper amplifiers serves as both a post amplifier and output buffer, Fig. 4a. Each individual gain stage, Fig. 4b, consists of a CMOS inverter followed by an inverter with resistive feedback. The first inverter acts as a transconductance element which converts the input small-signal voltage to a small-signal current that is then amplified and converted to an output voltage through the second inverter by way of resistive feedback. This second stage is effectively another TIA and extends the BW of the inverter cascade by reducing the impedance seen at the output node, reducing gain. By then further cascading amplifier unit cells, the desired total gain can be achieved without significantly sacrificing BW.

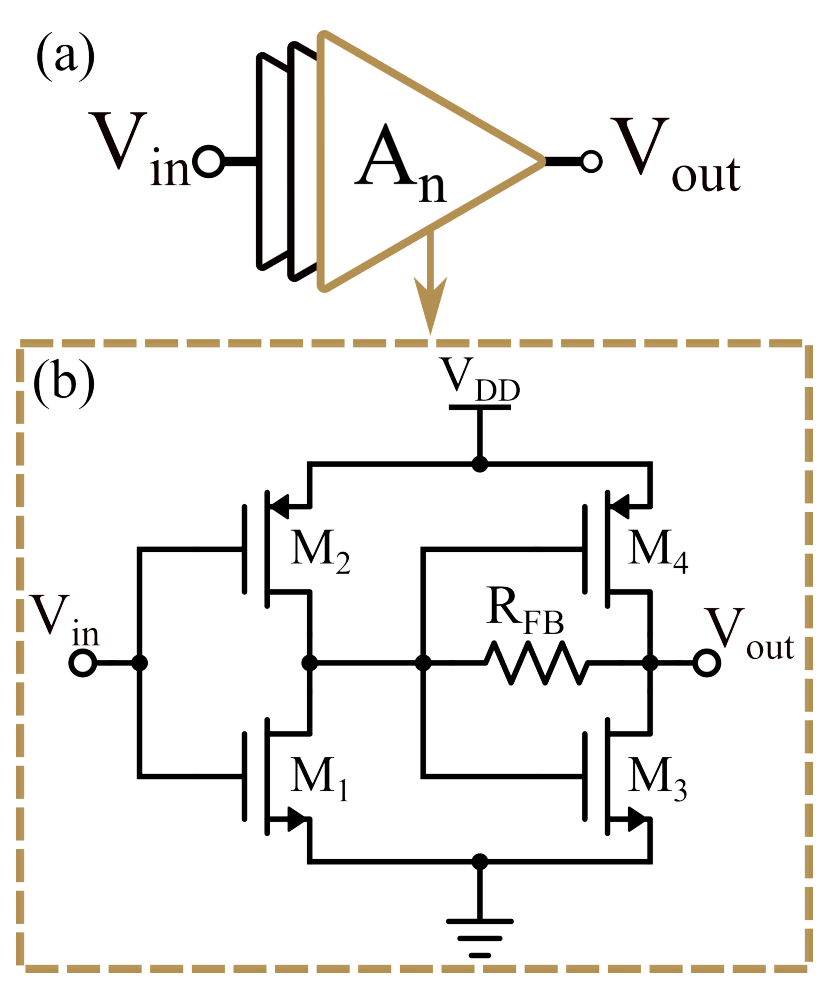

**FIGURE 4.** Block diagram of cascaded Cherry-Hooper amplifier-based post amplifier and OB architecture (a) with single gain stage schematic (b).

The voltage gain transfer function of each stage can be computed and simplified (assuming small-signal MOSFET output resistances are negligible) to:

$$\mathrm{A_v}(s) = \frac{g_{m1}+g_{m2}}{g_{m3}+g_{m4}} \times \frac{(g_{m3}+g_{m4})\mathrm{R_{FB}}-1}{1+s\dfrac{C_1+C_2}{g_{m3}+g_{m4}}+s^2\dfrac{\mathrm{R_{FB}}C_1C_2}{g_{m3}+g_{m4}}} \tag{8}$$

Where $g_{mN}$ is the small-signal transconductances of $M_N$, $C_1$ is the total capacitance at the drain of $M_1$, and $C_2$ is the total capacitance at the output node. Note that for sufficiently large $R_{FB}$, the DC voltage gain can be reduced to:

$$\mathrm{A_v}(0) = (g_{m1}+g_{m2})\mathrm{R_{FB}} \tag{9}$$

Due to the second order LPF type response, a 3-dB bandwidth can also be computed; however, in practice it is more convenient to maximize the bandwidth given a fixed gain and power budget. Additionally, noise analyses for the post amplifier and OB are omitted because the TIA gain is sufficiently large to make the contribution for each amplifier stage negligible when referred to the input node of the AFE. For this design, a single stage gain of ~ 5 dB is sufficient to meet design targets, requiring three total stages. Note that the transistor sizing in the combined post amplifier and OB follows the same exponential horn structure referenced in Section II B.

### *D. DESIGN FINALIZATION*

After transistor sizing and resistor value selection to ensure target gain, bandwidth, noise, stability, and power consumption, the initial design was simulated. The frequency domain behavior of the AFE, Fig. 5, highlights the ideal response prior to layout dependent effects. Note how the slight gain peaking in the TIA core extends the lesser BW of the post amplifier and OB to achieve a near Butterworth total transimpedance gain response. The initial simulations demonstrate 63 dBΩ transimpedance gain with an analog bandwidth of 30 GHz.

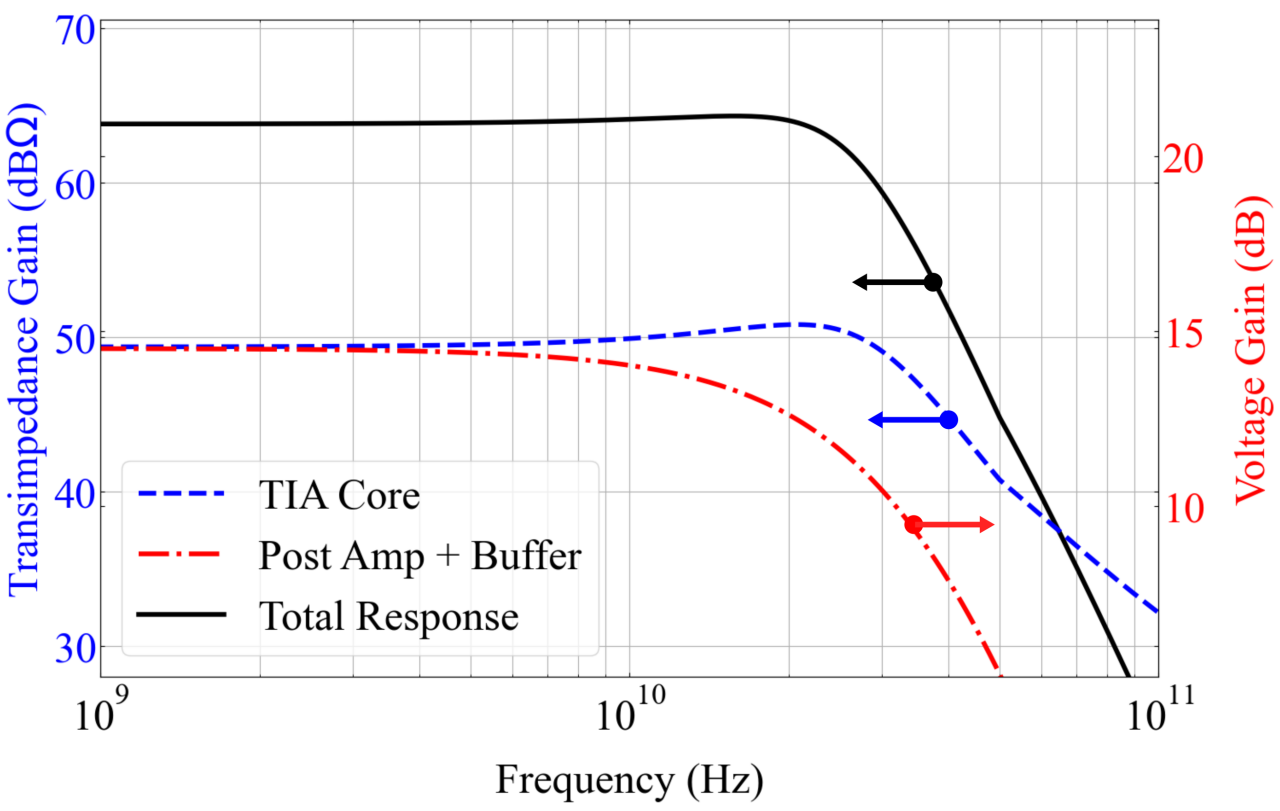

**FIGURE 5.** Schematic simulated transfer functions for various stages of the AFE including the TIA core, post amplifier and buffer, and the total response.

Because the AFE is inverter-based and DC-coupled, process mismatch and photodiode dark current can shift internal operating points. A slow DC offset cancellation loop can be added around the TIA or post amplifier chain to restore the nominal common-mode voltage without affecting the high-speed signal path. Similarly, tuning the feedback and source-degeneration resistors provides variable gain amplification which can be done by implementing them as triode region devices, with the gate bias controlled by an external automatic gain control feedback loop. Since these loops operate at frequencies far below the target baud rates, they improve robustness across process, voltage, temperature, and input optical power without degrading the broadband response. Conversely, these additions may substantially increase total power consumption and optimizing this is the subject of future work.

## III. PHYSICAL DESIGN

Although circuit and system level considerations are critical in the initial phases of AFE design, the physical implementation must be considered to anticipate and leverage layout parasitics for optimal measured performance. We utilize a methodical co-optimization approach for sizing transistors and selecting component values.

The $g_m$-over-$I_D$ methodology [17] was employed to determine preliminary transistor gate widths (W). Once a fixed width has been chosen for a given device, layout optimization and simulation feedback are performed. Layout and routing parasitics, primarily stray capacitance and routing resistance, greatly decrease overall system performance. These RC chains form cascaded low-pass filters which may reduce the 3-dB BW of an amplifier to well below the BW designed for. Each transistor was laid out with several parallel multi-finger devices, reducing effective gate resistance [18,19], to combat these challenges.

As an example, four parallel eight 1 µm-finger NMOS devices were utilized to form a transistor with an effective width of 32 µm, Fig. 6.

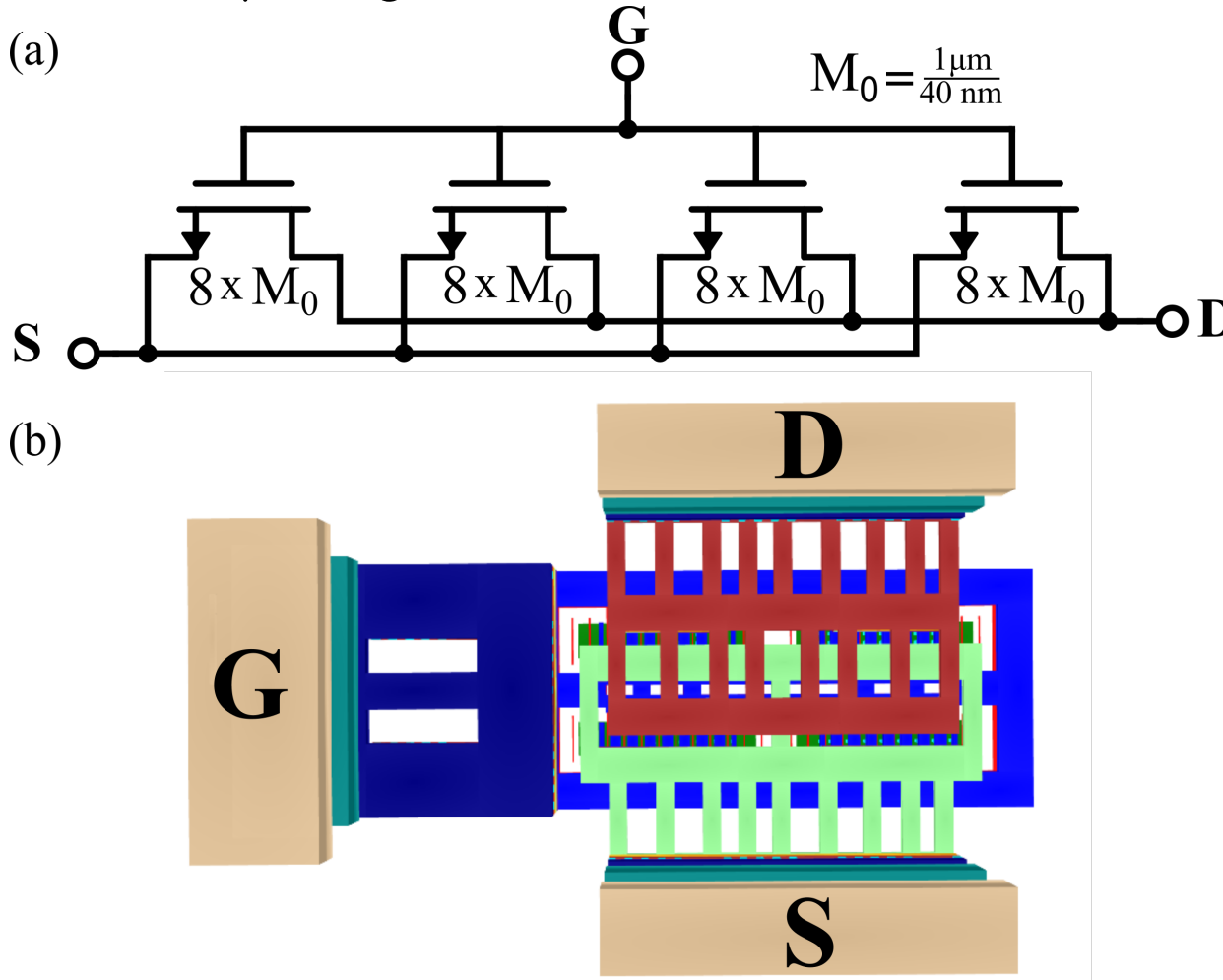

FIGURE 6. Schematic of 32 µm wide transistor consisting of 4 parallel transistors each with eight, 1 µm wide fingers (a) and nominal top-down metallization structure (b).

Note that this implementation includes routing from the bottom metal layer to the top metal layer for interstage component routing. After the initial transistor-unit cell was designed, the layout parasitics were extracted using a full-wave 3D simulation tool [20] to create an effective model of the physical device. Using this model, the component values and transistors widths were iteratively optimized to yield the desired performance.

The full layout, including ground planes, decoupling capacitors, and signal pads, is simulated using the same full-wave 3D simulation tool, Fig. 7. The final, worst-case (maximum parasitic resistance and capacitance models) post-layout simulated design achieved a low-frequency transimpedance gain of 61.7 dBΩ, a 3-dB BW of 28.9 GHz, and a minimum IRNC PSD of 7.4 pA/Hz$^{1/2}$. This simulation result includes the finite BW, capacitance, and contact resistance of the monolithic photodiode on the input. The total footprint (including pads and decoupling capacitors) is 650 µm x 750 µm, while the core footprint was only 20 µm x 135 µm.

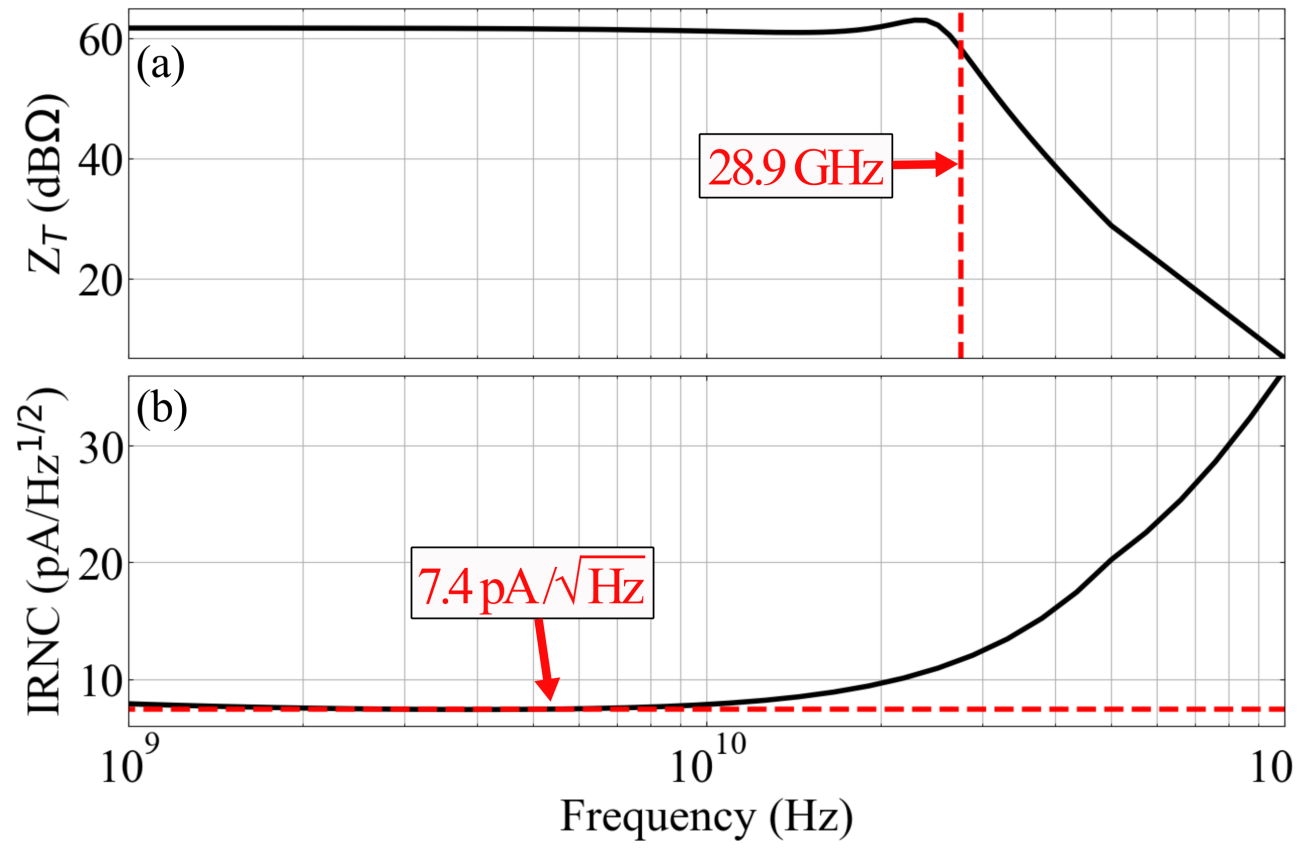

FIGURE 7. Post-layout simulated AFE performance spectra including the transimpedance gain (a) and the IRNC (b) accounting for finite PD BW and layout parasitics.

## IV. EXPERIMENTAL RESULTS

Two variants of the final AFE design were taped out on the GlobalFoundries Fotonix™ platform, one with the input connected to a photodiode and the other with RF pads at the input. First, the DC operation and power consumption were analyzed. Following this, the AC response is measured and compared to the post-layout simulations. Next, the noise behavior of the circuit was investigated. Finally, a time-domain study was performed to demonstrate functionality of the receiver front end for various baud rates of on-off keyed non-return-to-zero (OOK-NRZ) and four level pulse-amplitude modulation (PAM-4) signals. The basic experimental setup, Fig. 8, centers around the device-under-test (DUT) which was fixed in place to the top of a probe station.

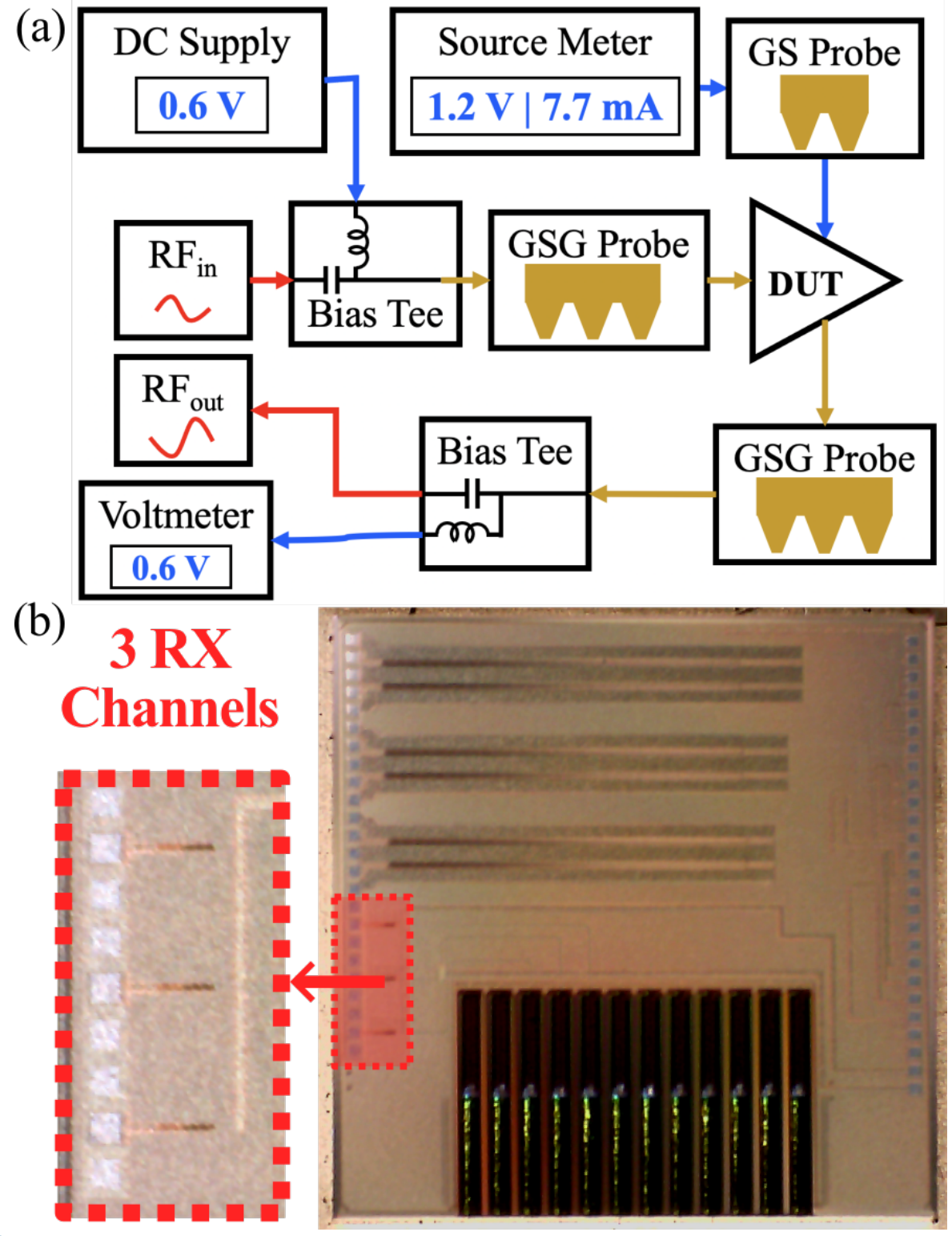

FIGURE 8. Block diagram of experimental setup for electrical characterization of the optical receiver (a) and microscope image of the fabricated die with three devices-under-test (b).

The supply voltage was sourced through a GS probe from a source meter to record the current draw, while the RF input and output were coupled through bias tees to a 40 GHz BW GSG probes. For AC response measurements, the RF input and output signals were sourced and detected by the Agilent 8510C vector network analyzer (VNA). For time domain measurements, the input RF signal was sourced by the Keysight M8198A arbitrary waveform generator, while the output signal was coupled to a 60 GHz BW Infinium DCA-X 86100D sampling oscilloscope. For both high speed experiments, the DC input voltage was set to $V_{DD}/2$ and the output DC voltage was monitored using an external voltmeter.

### A. DC OPERATION

To ensure low-power operation, the AFE current and power consumption were measured as a function of the supply voltage, Fig. 9. The receiver was designed with transistors that utilize a nominal supply voltage of 1 V. However, due to the symmetric nature of the CMOS inverter amplifiers within the TIA, post amplifier and OB, the system can be powered with a $V_{DD}$ ranging from 0.8 V up to 1.2 V. For both the time domain and noise experiments, the system was powered with the 1.2 V. At this operating point, the DC current consumption was measured to be 7.68 mA consuming only 9.22 mW of power. For specific data constrained applications requiring even less power consumption, the device can be powered with a 0.9 V supply consuming 2.5 mW, while still supporting 32 GBaud signaling schemes.

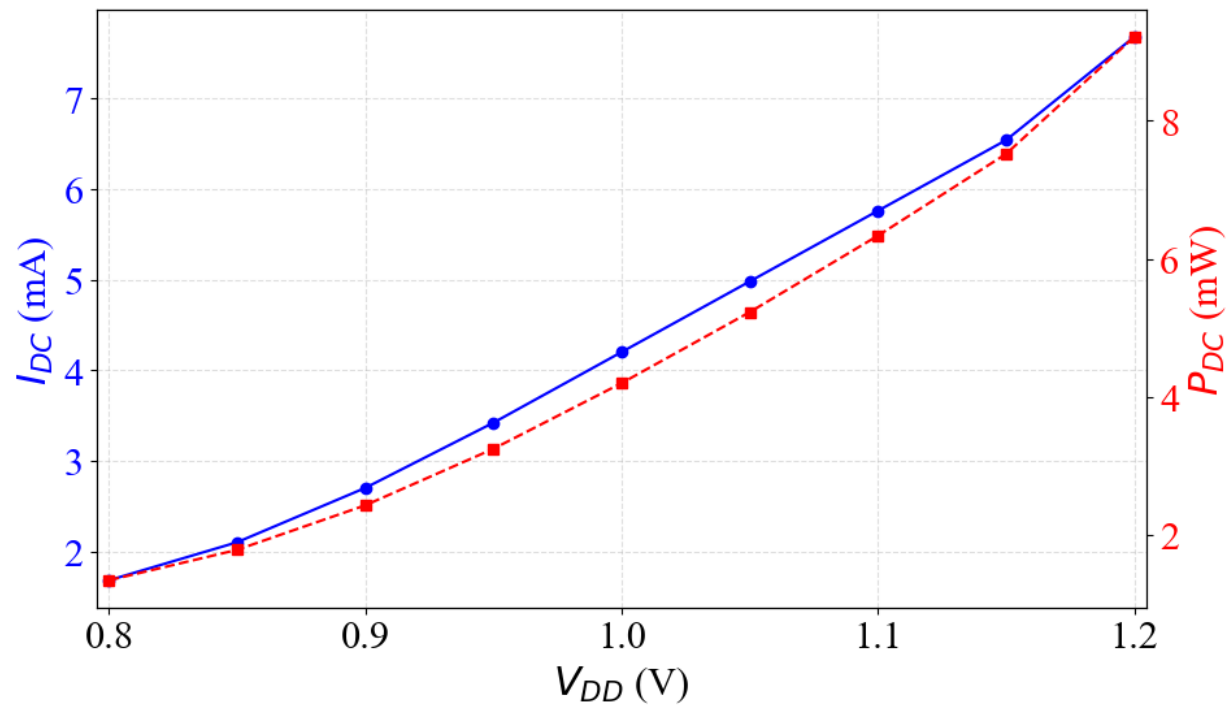

**FIGURE 9.** Measured DC current (left) and computed DC power consumption (right) as a function of supply voltage.

### B. AC OPERATION

AC operation was validated by measuring the magnitude response of the transimpedance gain. The VNA calibration was performed such that the measurement reference plane included only the RF pads and the DUT. This response is shown alongside the post-layout simulated response, Fig. 10. The measured and simulated results reveal strong agreement in the low frequency regime, with the measured (simulated) gain being 61.5 dBΩ (61.7 dBΩ). The measured BW is slightly higher at 29.5 GHz compared to the simulated 28.9 GHz due to the lack of the 40 GHz LPF photodiode response. Additionally, the response rolls after faster than simulated, likely due to more uncertainty in the calibration at high frequencies.

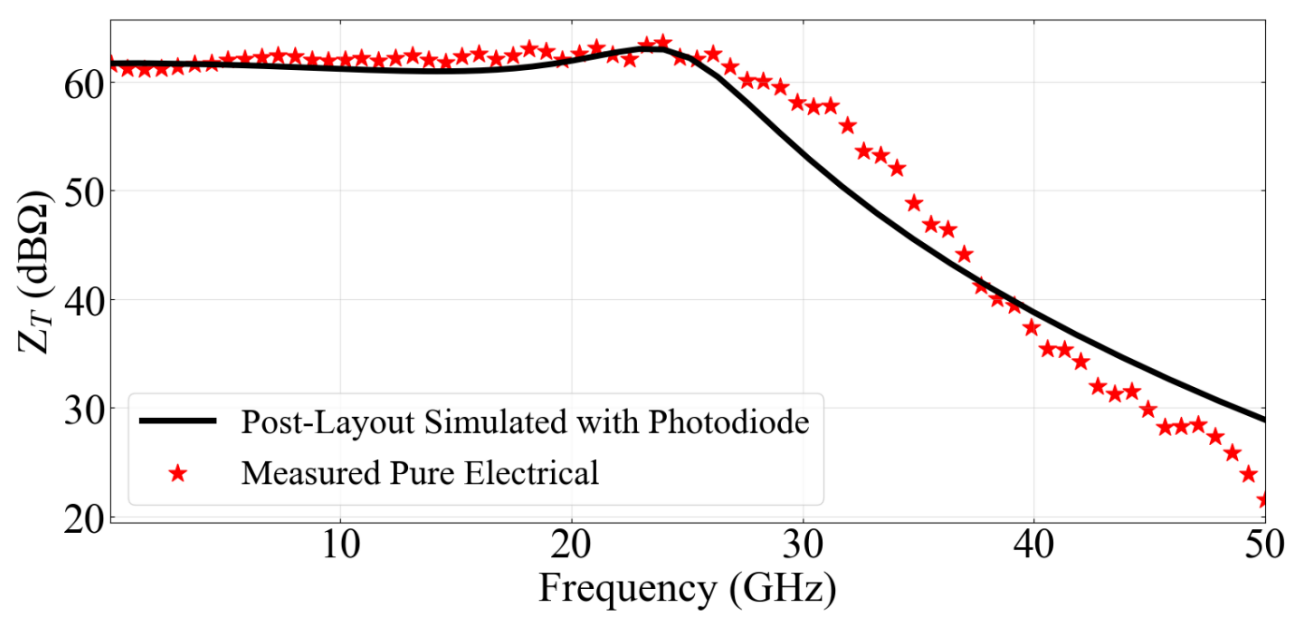

**FIGURE 10.** Filtered measured pure electrical magnitude response of the transimpedance gain (red) overlaid on the post-layout simulated magnitude response of the AFE including the finite photodiode BW (black).

### C. NOISE

The noise behavior of the AFE was determined by measuring the output voltage of the TIA with no input current. To include the impact of the photodiode noise on the total AFE noise, the receiver with the optical input was measured. A histogram of the output voltage was taken and fitted to a Gaussian distribution, Fig. 11a. The mean was measured as 2.05 mV, while the RMS standard deviation was measured as 896 μV RMS, corresponding to an input referred noise current of 737 nA RMS. The receiver sensitivity ($P_{sense}$) was computed as a function of desired Q-factor for an OOK-NRZ signal, Fig. 10b, from the relationship [21]:

$$P_{\text{sense}} = \frac{2Qi_n^{RMS}}{R} \tag{10}$$

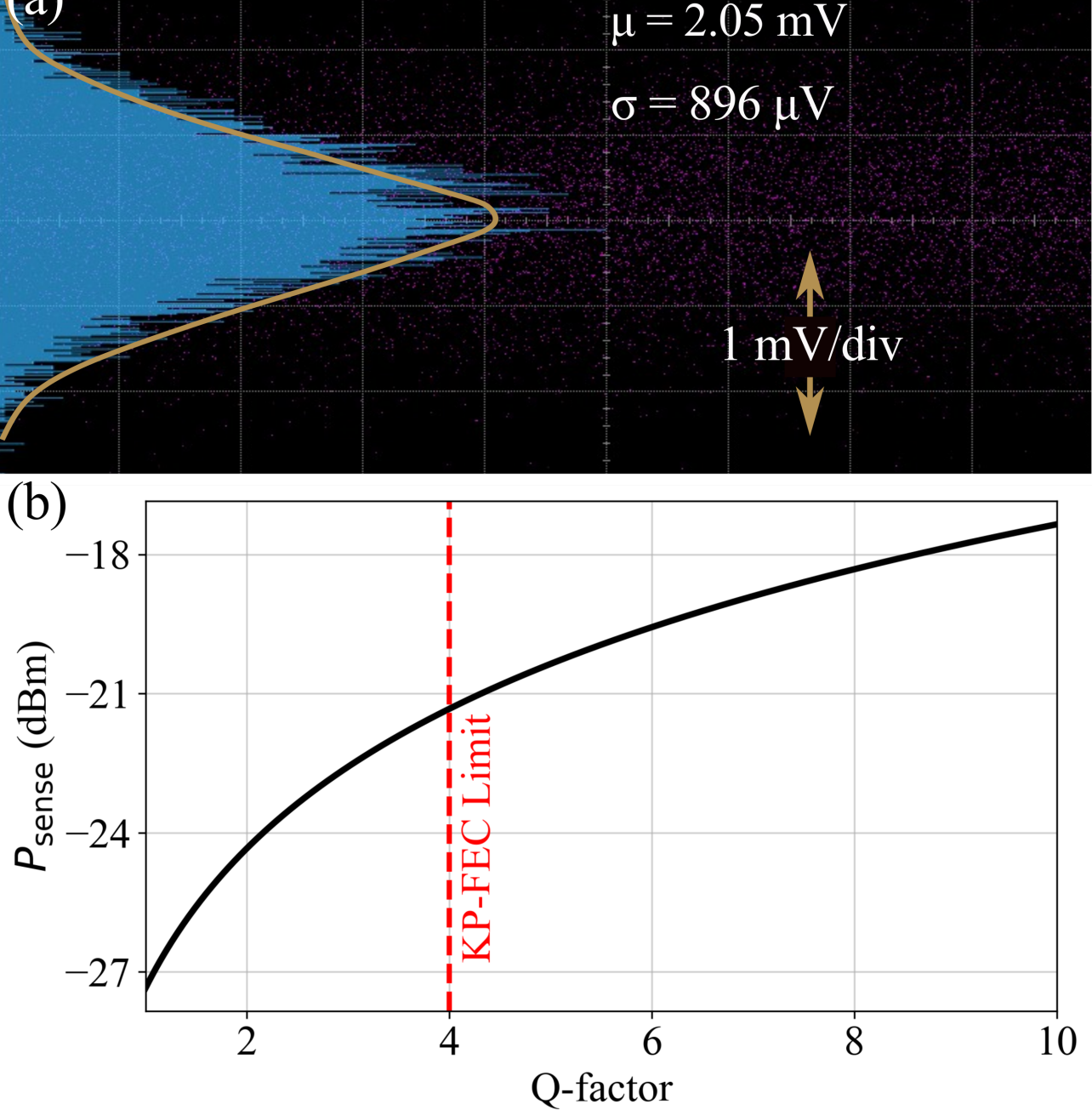

**FIGURE 11.** Measured output noise voltage histogram (a) and the computed receiver sensitivity as a function of Q for an OOK-NRZ signaling scheme (b).

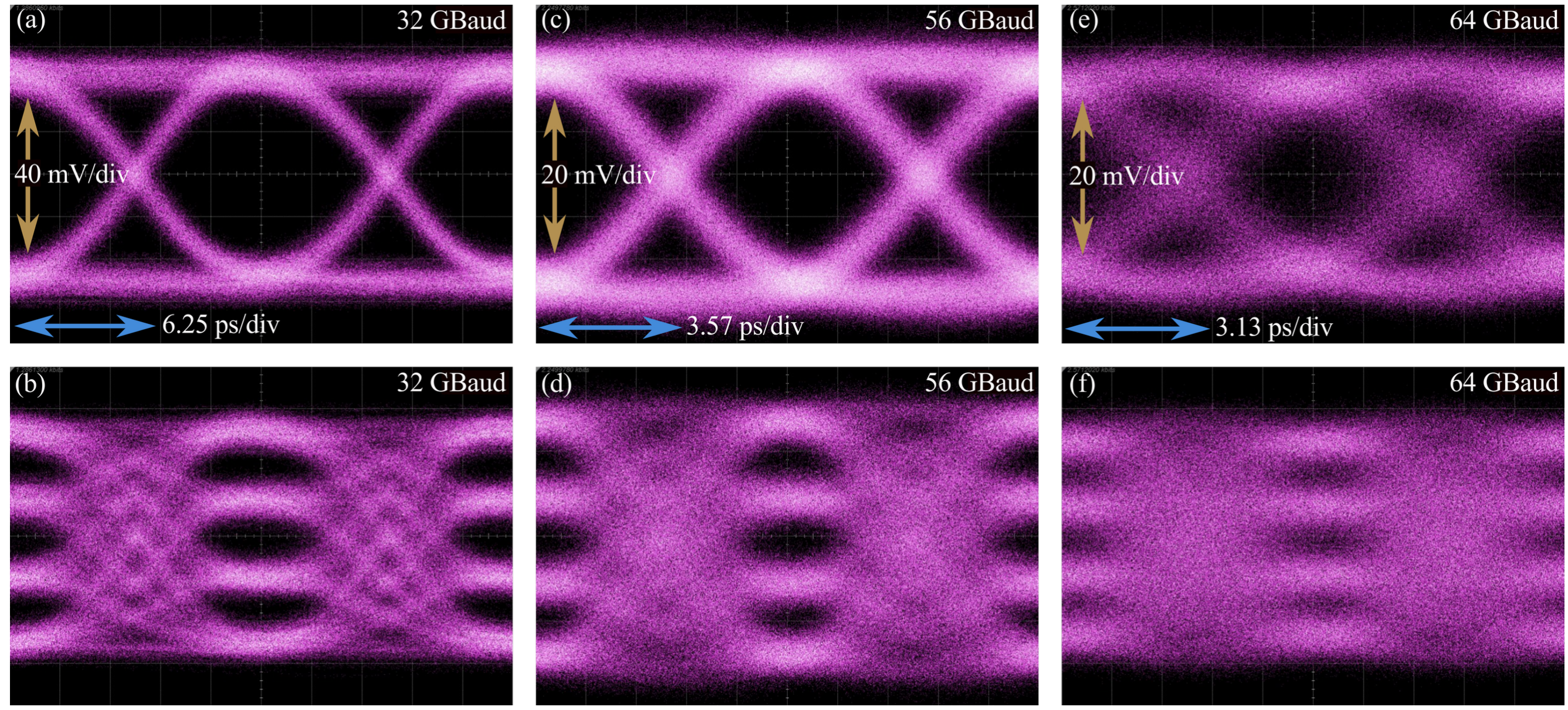

**FIGURE 12.** Measured eye diagrams at the output of the optical receiver with de-embedding of all external circuitry excluding RF probes and the DUT. Columns from left to right correspond to 32 GBaud (a,b), 56 GBaud (c,d), and 64 GBaud (e,f) respectively, while the top (a,c,e) and bottom (b,d,f) rows are NRZ-OOK and PAM-4 respectively.

where Q is the desired Q-factor, $i_n^{RMS}$ is the RMS IRNC, and R is the photodiode responsivity. Note (10), assumes the noise associated with both a logical zero and one are identical. It also neglects eye width reduction as baud rates increase and is thus a minimum value. We compute a minimum sensitivty of -21.3 dBm with standard forward error correction (KP-FEC) technqiues and -19.57 dBm without FEC.

### D. TIME-DOMAIN BEHAVIOR

Finally, the AFE was validated by a full electrical eye diagram measurement. The input signal was varied between OOK-NRZ and PAM-4 modulation formats while the baud rate was swept from 32-64 GBaud for each modulation format. The recorded eye diagrams demonstrate high-performance out to 64 GBaud, Fig. 12. Additionally, the RF cables, connectors, and bias tee responses were de-embedded, highlighting the pure GSG probe and DUT response. The 32 GBaud and 56 GBaud eye diagrams yield error free performance when accounting for forward error correction (FEC). For the 64 GBaud PAM-4, the individual levels are discernable but require additional equalization for error-free operation. Therefore, we determine the maximum bit rate to be 112 Gbps with the 56 GBaud PAM-4 signaling, corresponding to 0.08 pJ/bit. Additionally, the linearity of each PAM-4 eye has been characterized by their respective ratio level mismatch (RLM) defined as:

$$\mathrm{RLM} = \frac{3min(\Delta_1, \Delta_2, \Delta_3)}{\Delta_1 + \Delta_2 + \Delta_3} \tag{11}$$

where $\Delta_n$ is the voltage difference between the nth and n-1th level of the PAM-4 eye. The measured RLM values of the 32 GBaud, 56 GBaud, and 64 GBaud eyes were 0.945, 0.951, and 0.942 respectively. These results, demonstrate sufficient linearity for short reach applications operating at 56 GBaud.

### E. COMPARISON WITH REPORTED WORK

Many optical receiver analog front ends have been proposed and discussed in the literature, Table 1. The design presented here achieves the lowest energy per bit of any receiver reviewed at 0.08 pJ/bit. Additionally, this design attains the second lowest sensitivity, requiring only -19.57 dBm input optical power for a bit error ratio of $10^{-12}$. Although this metric is for the ideal case, in raw BW and transimpedance gain, this design performs alongside the best reported with a greater energy efficiency. This is accomplished by designing for a modest transimpedance gain of 61.5 dBΩ, but due to the low IRNC the overall receiver sensitivity is comparable to the best reviewed. Additionally, the die area of the AFE is smaller than most designs reviewed enabling ultra-dense system integration. Although the SiGe BiCMOS receivers reviewed achieve comparatively large BWs [7,22] that are challenging for conventional CMOS platforms to match, packaging parasitics resulting from hybrid integration limit final system level data-rates and noise performance. Therefore, monolithic CMOS proves to be a good choice for short reach applications.

## V. CONCLUSION

This work presents a compact, low-noise and low-power monolithic optical receiver analog front end that requires less than 0.08 pJ/bit. Our modified AVCF architecture was shown to be an ideal topology for implementing a low power TIA in a monolithic platform. Together with a multistage inverter-based Cherry-Hooper post amplifier and output buffer, the amplifier cascade forms an AFE that achieves record energy

TABLE I
PERFORMANCE METRICS FOR REPORTED OPTICAL RECEIVERS IN THE LITERATURE

| References | [12] (2021) | [13] (2024) | [22] (2024) | [23] (2024) | [24] (2025) | [7] (2025) | **This Work** |
|---|---|---|---|---|---|---|---|
| Process | 40 nm CMOS | 65 nm CMOS | 130 nm SiGe BiCMOS | 28 nm CMOS | 28 nm CMOS | 55 nm SiGe BiCMOS | **40 nm CMOS** |
| $f_t$ (GHz) | 250[a] | 160[a] | 260 | 300[a] | 300[a] | 320[a] | **280** |
| TIA Architecture | AVCF | AVCF | Shunt Feedback | Multistage Inverter | Multistage Inverter | Travelling Wave | **AVCF** |
| Supply Voltage (V) | 1 | 1.2 | 3.3 | 1/0.7 | 1.05 | 2.5 | **1.2** |
| $Z_T$ (dBΩ) | 63.8 | 70.7 | 74 | 82 | 92 | 77 | **61.5** |
| BW (GHz) | 24.3 | 16.3 | 40 | 7 | 5.3 | 64 | **29.5** |
| IRNC (pA/Hz$^{1/2}$) | 95.6[a] | - | - | 11 | 3.6 | 18.8 | **7.4** |
| Data Rate (Gbit/s) | 30 | 25 | 112 | 12.5 | 10 | 224[c] | **112** |
| Sensitivity (dBm), BER = $10^{-12}$ | -5.84[b] | - | -9.7 | -10.7 | -20.7 | -5.9 | **-19.57[b]** |
| Area (mm$^2$) | 0.63 | 0.45 | 0.816 | 0.0007[d] | 0.6 | 1.125 | **0.44** |
| Power (mW) | 37.5 | 22.6 | 180 | 1.38 | 121 | 487.5 | **9.2** |
| Energy Efficiency (pJ/bit) | 1.25 | 0.90 | 1.61 | 0.11 | 12.1 | 2.18 | **0.08** |

[a] estimated from available data, [b] computed from IRNC, [c] including channel equalization, [d] core area not including pads or decoupling capacitors

per bit compared to similar designs reported in the literature. Moreover, a systematic layout aware transistor sizing scheme was introduced to maximize post-fabricated system performance, leveraging the inherent routing resistance and capacitances. Ongoing work includes demonstration of a full transceiver system, including a similarly designed optical transmitter. Additionally, methods for reducing transceiver footprint in parallel with power consumption are being explored [25]. This work demonstrates the utility of monolithic electronic-photonic optical transceiver systems for the SWaP-C constrained AI datacenter and aerospace platforms of the future.

## ACKNOWLEDGMENT

This material is based upon work supported by the National Science Foundation Graduate Research Fellowship under Grant No. DGE-2039655 and National Science Foundation "EPICA", Grant No. 2052808.

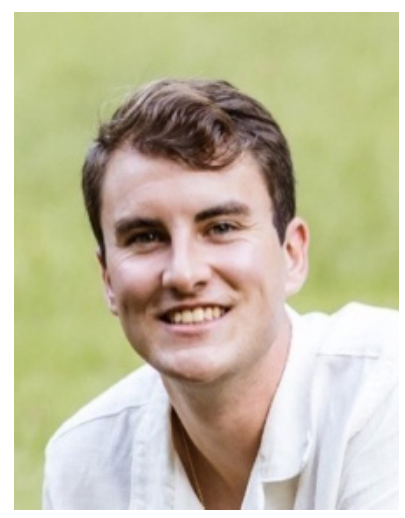

**ROBERT P. PESCH** (M'24) received the B.S. (2024) and M.S. (2025) degrees in electrical engineering from the Georgia Institute of Technology, Atlanta, USA. He is currently a graduate researcher pursuing his Ph.D. in Electrical and Computer Engineering at the Georgia Institute of Technology under the advisement of Dr. Stephen Ralph. His research interests include the design of ultra-wideband CMOS optical transceivers and RF and microwave photonic systems. Robert is a current NSF graduate research fellow and IEEE graduate student member.

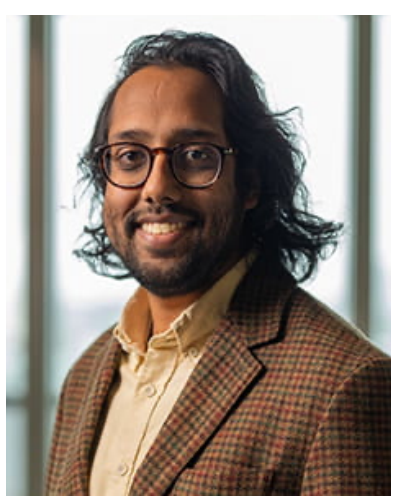

**ARJUN KHURANA** (M'21) was born in Plano, Texas. He graduated with a Bachelor of Science in Computer Science and a minor in Physics from the University of Texas at Dallas in 2021. Since then, he has been a Ph.D. student working with Dr. Stephen Ralph specializing in optical 3D heterogeneous integration and novel VCESL architectures. This is strongly backed by his extensive internship experiences at Dallas Quantum Devices and Finisar Corporation. He has authored several publications relating to photonics inverse design and novel VCSEL structures.

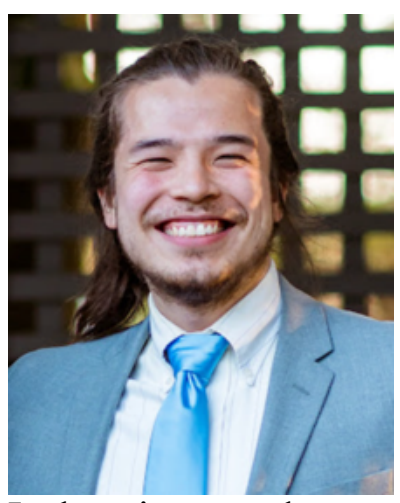

**JOSHUA J. WONG** received the B.S. degree in Computer Engineering from the Georgia Institute of Technology in spring 2023 and received his M.S. degree in Electrical and Computer Engineering at the Georgia Institute of Technology in fall 2024. He is currently pursuing in Ph.D. in ECE at Georgia Tech under the advisement of Dr. Stephen E. Ralph. His research interests include integrated photonics, inverse design, solid-state physics, and electromagnetics. Joshua is a graduate student member of the IEEE and a recipient of the National Defense Science and Engineering Graduate (NDSEG) Fellowship.

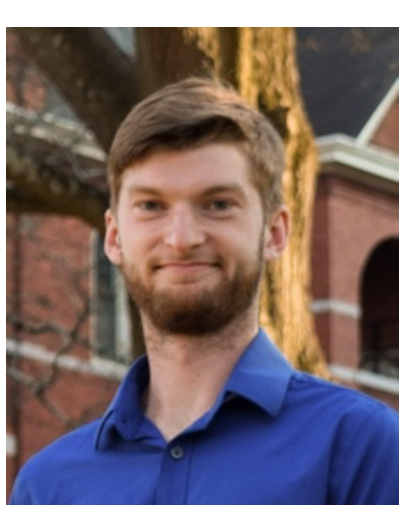

**JOEL B. SLABY** was born in Pittsburgh, Pennsylvania. He received his master's and bachelor's degree in electrical engineering from Georgia Institute of Technology in 2022 and 2021, respectively, where he is currently pursuing a Ph.D. in electrical engineering. He has been a member of Dr. Stephen Ralph's ultrafast optical communications laboratory for six years and is currently developing novel methods for designing foundry compatible integrated silicon photonic components and systems. He is investigating and designing integrated circuits robust to harsh environments through the NSF IUCRC Electronic and Photonic Integrated Circuits for Aerospace (EPICA). He has co-authored several journals related to density-based topology optimization and presented at several conferences for similar work.

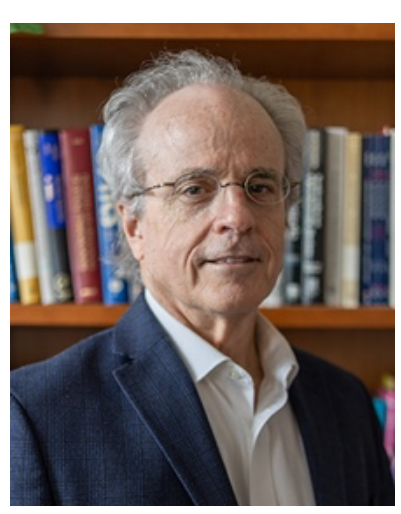

**STEPHEN E. RALPH** (Senior Member) received the B.E.E. degree (Hons.) in electrical engineering from Georgia Institute of Technology (Georgia Tech), Atlanta, GA, USA, in 1980, and the Ph.D. degree in electrical engineering from Cornell University, Ithaca, NY, USA, in 1998. His research focused on the optical detection of highly nonequilibrium transport in heterojunction devices. In 1988, he began a postdoctoral position at AT&T Bell Laboratories. In 1990, he joined the IBM T. J. Watson Research Center, Yorktown Heights, NY, USA. In 1992, he joined as a Faculty Member with the Physics Department, Emory University, Atlanta. In 1998, he was an Associate Professor of electrical and computer engineering with Georgia Tech, where his work currently focuses on the development of ultrafast optical devices for telecommunications. He is a member of American Physical Society, the IEEE Lasers and Electro-Optics Society, and the Optical Society of America.